# MODELLING RF POWER AMPLIFIER TO STUDY ITS NON LINEAR EFFECTS ON RF COMMUNICATION SYSTEM, WITH BER AS A PERFORMANCE MEASURE.

Aripirala Manoj Kumar, G. Bharath Reddy, G. Krishna Chaitanya Reddy.

[1]Department of Electronic and Communication Engineering, IIIT-Allahabad.
manojkumar.aripirala@gmail.com
[2]Department of Electronic and Communication Engineering, IIIT-Allahabad.
bharathrgaddam@gmail.com
[3]Department of Electronic and Communication Engineering, IIIT-Allahabad.
chaitu.iiita@gmail.com

## ABSTRACT

*This paper is a study of non-linear effects of RF Amplifiers on Communication Systems Performance. High speed data communication is made possible by Multilevel Modulation schemes. This paper presents a study of these non linear effects on multilevel Modulation schemes like MPSK and MQAM. We make use of Bit Error Ratio (BER) as performance measure. BER vs SNR (Signal-to-Noise Ratio) curves provide comparison between the non linear effects caused by Gain Compression in particular.*

## 1. INTRODUCTION

In this section we discuss the need to study the non linear effects in RF Amplifiers. This is followed by the modelling of the Communication system in the $2^{nd}$ section. In the $3^{rd}$ section we describe the method to model the RF Amplifier from different parameters. In $4^{th}$ we discuss the results of applying the Amplifier model in the communication system simulator and conclude in the next section. This is followed by Acknowledgements and References.

The present day wireless communication applications demand high speed data transmissions. This is made possible by multilevel modulation schemes like M-ary Phase Shift Keying (MPSK) and M-ary Quadrature Amplitude Modulation (MQAM). The Non Linear Effects in RF Amplifier like Gain Compression cause Power Efficiency problems especially in multilevel modulation scheme causing a trade off. There are methods to compensate for these non linear effects.

We can increase the input power to the Amplifier but this will cause power efficiency losses. We can also use Compound Semiconductors like GaAs in the RF amplifiers but that will increase the cost of Amplifier making it unsuitable for low cost and small scale integrated circuit implementations. By studying these non linear effects we can optimise the trade offs.

## 2. COMMUNICATION SYSTEM SIMULATOR

The simulation has been done in MATLAB 7.7.0471. In the simulator the transmitter transmits bits at a given data rate. These bits are then modulated using multilevel modulation schemes. Then the modulated bits are transmitted through a channel with AWGN. They are then demodulated and sent to the receiver. The received bits are then compared with the sent bits and then BER is evaluated. This BER depends on the SNR of the signal. The following figure shows a block diagram of different elements in the simulator –

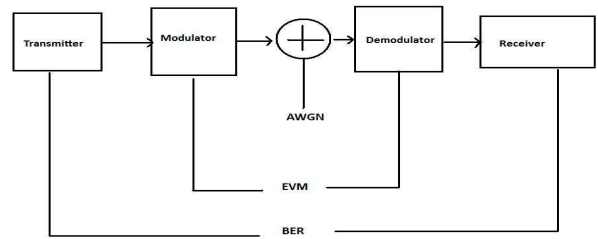

Fig.1: Block Diagram of Communication system simulator.

Apart from BER another parameter called Error Vector Magnitude (EVM) also serves as a good performance measure but we restrict our discussion to BER.

### 2.1. MPSK

Phase Shift Keying uses distinct phases and each of these phases represent a unique symbol pattern of bits. The figure below shows the scatter plot of BPSK signal transmitted at 30000 bits per second.

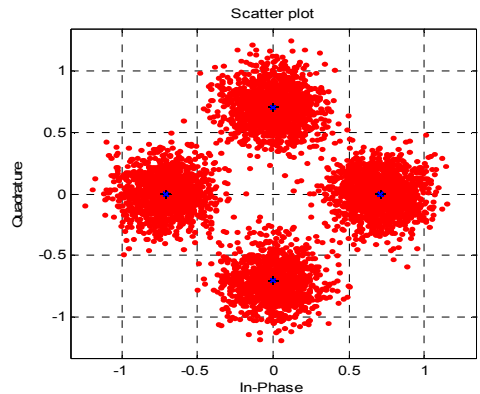

Fig.2: Scatter plot of 4-PSK .

The above figure shows a acatter plot of 4-PSK modulated signal. Scatter plot shows the In-Phase and Quadrature Components of the modulated signal. Generally after modulation the symbols will be complex( in x +iy for). If the signal values fall beyond a particular threshold in the scatterplot then it will be detected as another symbol thus causing errors. BER vs SNR curves of MPSK for different values of M is shown below –

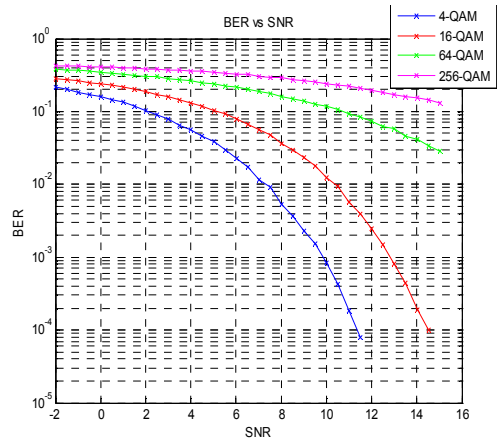
Fig.3: BER vs SNR plot of MPSK.

## 2.2. MQAM

Quadrature Amplitude Modulation is both a digital scheme as well as an analog scheme. It conveys two analog signals or digital bit streams using Amplitude shift keying and these are usually 90 degrees out of phase and so the name Quadrature. The following figure shows signal constellation and scatter plot of the MQAM signal transmitted at 30000 bits per second –

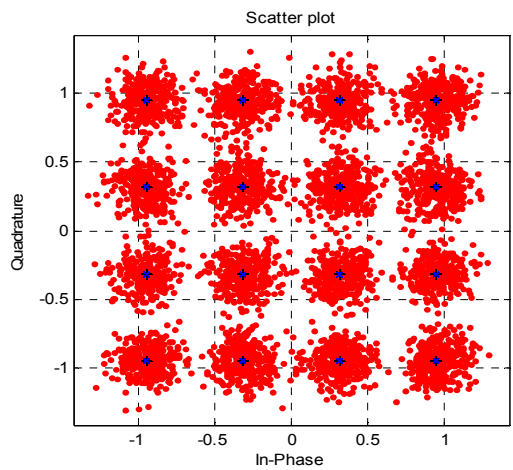
Fig.4: Scatter plot of 16-QAM.

The following plot shows the BER vs SNR curves of MQAM for different values of M.

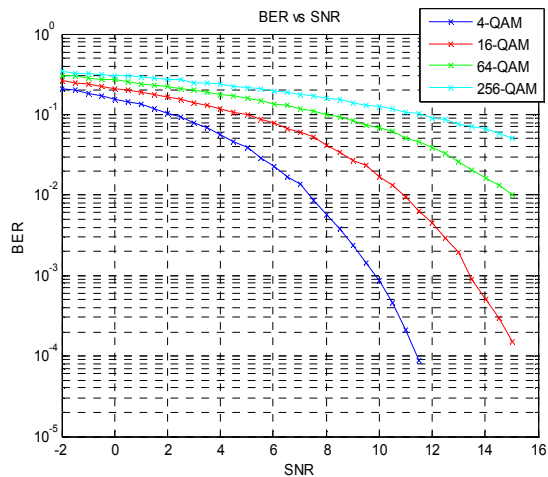
Fig.5: BER vs SNR plot of MQAM.

In the plots we can see that BER vs SNR is a decaying curve because as the SNR increases Noise decreases and so the number of errors tend to decrease. During the calculation of BER vs SNR values the number of bit errors become really less and cause some irregularities so we increase the number of simulated bits and there by the ratio remains the same.

Now a parametrizable RF Amplifier model is to be applied to the simulator and the above BER vs SNR curves are used for comparison.

## 3. AMPLIFIER MODEL

The RF Amplifier must be modelled in such a way that it exhibits Gain Compression Effect. And the amount of compression and point of saturation is to be determined by the given parameters.

### 3.1. Gain Compression

In RF Amplifiers the Gain is not linear. At higher input levels the gain tends to saturate. This effect is called Gain Compression and the parameter to signify Gain Compression is 1-db Compression Point. It is defined as the input power at which the output is 1-db less than the expected value. The system equation of linear power Amplifier is given as –

$$y(t) = a_1 * x(t) \dots\dots\dots\dots\dots\dots\dots(1)$$

Because of the non linearities the system equation of the RF Amplifiers can be generalised as below –

$$y(t) = a_1 * x(t) + a_2 * x^2(t) + a_3 * x^3(t) \dots\dots\dots\dots\dots(2)$$

If we apply a sinusoid as the input to the above non linear system the resultant equation would be –

$$x(t) = A * \cos(\omega t)$$

$$y(t) = a_1(A\cos(\omega t)) + a_2(A\cos(\omega t))^2 + a_3(A\cos(\omega t))^3$$

$$= a_1(A\cos(\omega t)) + (a_2 A^2/2)(1 + \cos 2\omega t)$$
$$\quad + (a_3 A^3/4)(3\cos \omega t + \cos 3\omega t)$$

$$= a_2 A^2/2 + (a_1 A + 3 A^3/4) \cos \omega t + (a_2 A^2 \cos 2\omega t)/2$$
$$\quad + (a_3 A^3 \cos \omega t)/4$$

From the above equations we can observe that even order harmonics vanish if the system has odd symmetry. The nth harmonic grows approximately proportional to $A^n$. In the above equations we can see that the 1st harmonic depends on the coefficients a1 and a3. In modelling the amplifier we have to find out the values of a1 and a3 from given parameters.

In this paper we use Gain (G), output power at i-db compression point (P_1db) and saturated output power (P_sat) to find the coefficients. The following equations show how the values of a1 and a3 are derived from the above mentioned parameters –

Gain, $G = V_{out}/V_{in}$

Gain in db, $G\_db = 20*log(V_{out}/V_{in})$

In case of linear amplifier $V_{out}$ = y(t) and $V_{in}$ = x(t). And for a linear amplifier y(t) = $a_1$*x(t) as in eq(1). Substituting in above Gain equation we get –

$G\_db = 20*log(y(t)/x(t))$

$G\_db = 20*log(a_1*x(t)/x(t))$

$a1 = 10^{(G\_db/20)}$

From the above equation we get the coefficient of first order term in the non linear system equation. And since we are concerned about the first major harmonic we can neglect the second order term's coefficient. And so the system equation becomes –

$$y(t) = a_1*x(t) + a_3*x^3(t)\ldots\ldots\ldots\ldots(3)$$

So we need the third order term's coefficient a3. And the parameter P_1db is used to find a3. P_1db is the output power at 1-db compression point and so the output power is 1_db less than linear value. So –

$P\_1db + 1 = 20*log(a_1*x(t))$

$x(t) = 10^{(P\_1db + 1)/20}/a_1$.

From this we obtain the input voltage at 1-db compression. And the output voltage is given by –

$y(t) = 10^{(P\_1db/20)}$

Substituting in eq(3). we get –

$$a_3 = (y(t) - a_1*x(t)) / x^3(t)$$

So, for any given parameters we can successfully model the RF Amplifier. We modelled the Amplifier as a C++ class as a Matlab Executable File (.mex). The parameters are taken from Hittit Electronics and the specifications used belong to the HMC413QS16G/ HMC413QS16GE. The following are the specifications of the IC –

*Gain, G_db = 23db*
*Saturated Output Power, P_sat = 29.5dbm*
*Output Power at 1db Compression point, P_1db = 27dbm*
*Supply Voltage = +2.75V to +5V*
*Output power at $3^{rd}$ Interception Point, P_IP3 = 40dbm*

Using the above specifications the amplifier is modelled. The following figures shows the plot of Output Power to Input Power in db and Output Voltage to Input Voltage in volts of the Amplifier –

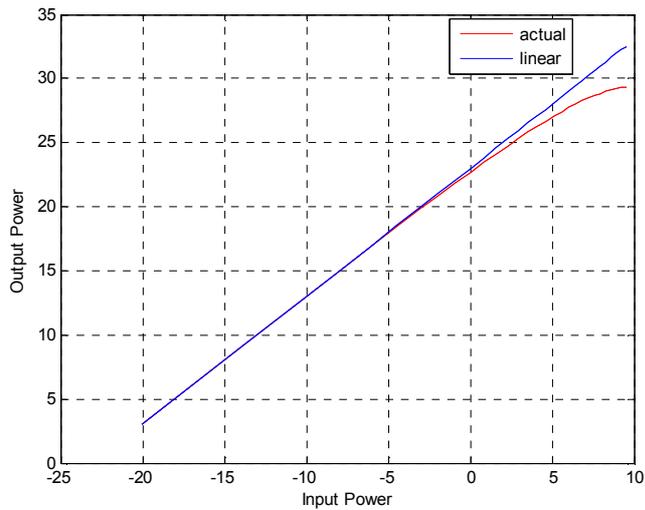

Fig.6: Output Power Vs Input Power (in db)

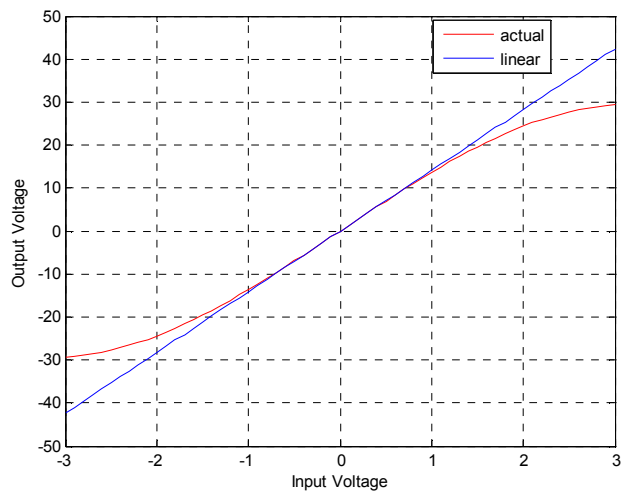

Fig.7: Output Voltage Vs Input Voltage

In the following section we discuss the results of applying the Amplifier Model to the communication system simulator.

## 4. RESULTS

The following figure shows the BER vs SNR curves of a 16-QAM modulated signal after applying the Amplifier Model. From the plot we can see that initially because of the gain that signal power increases. As a result BER value decreases. Then as the input signal energy increases the non linearitites come into effect increasing the number of errors and thereby increasing the BER. If the signal energy comes into this range of non linearities there will be distortions causing Amplifier Back off.

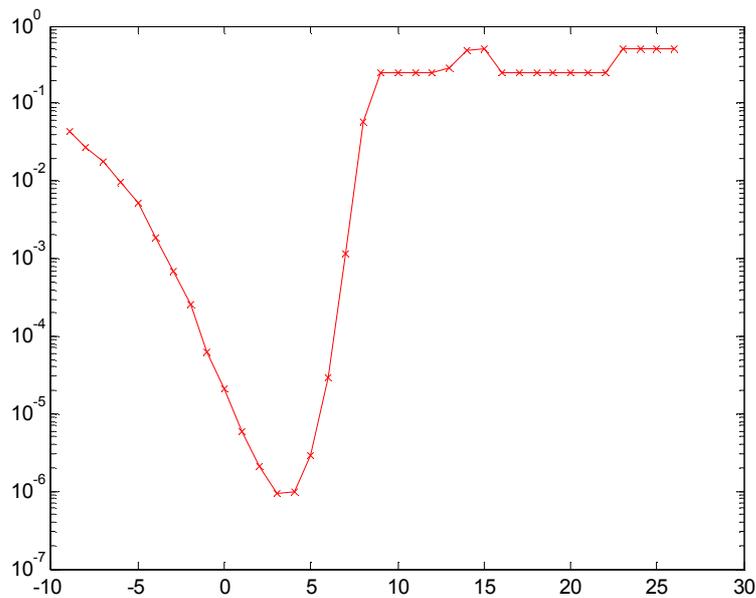

Fig.8: BER vs SNR curve of a 16-QAM modulated signal along with the Amplifier model.

These are the scatter plots that we get after applying the amplifier model to the communication system.

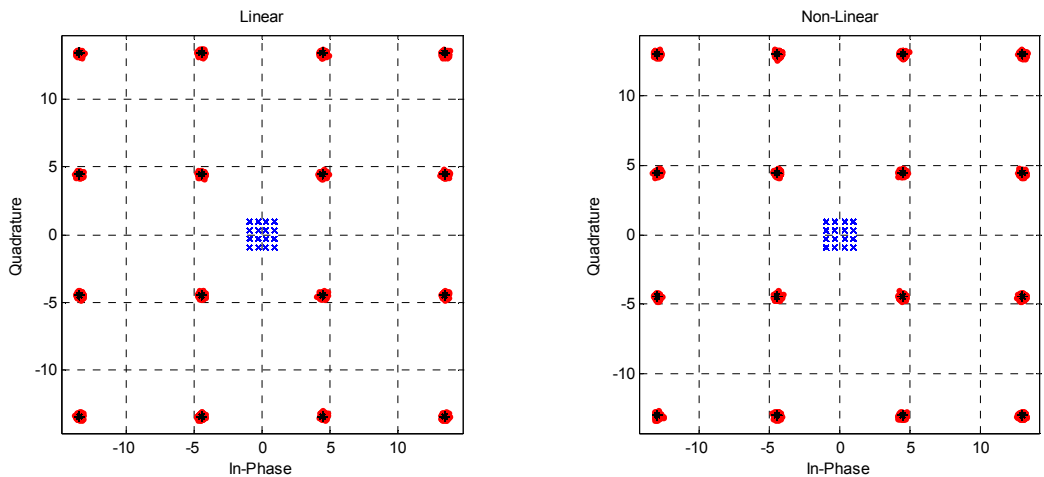

Fig.9: Scatter plots of 16-QAM signal for linear and non linear amplification.

The above figure shows the scatter plot of linear and non linear amplification. The scatter plot in blue is the scatter plot of actual 16-QAM modulated signal with average energy 1 and all the symbols with amplitude less than 1. Gain Compression causes errors in both Amplitude and Phase (because the symbol generally of the form x + iy and x and y are separately amplified). The following figure clearly shows the difference in Linear and Non linear behaviour.

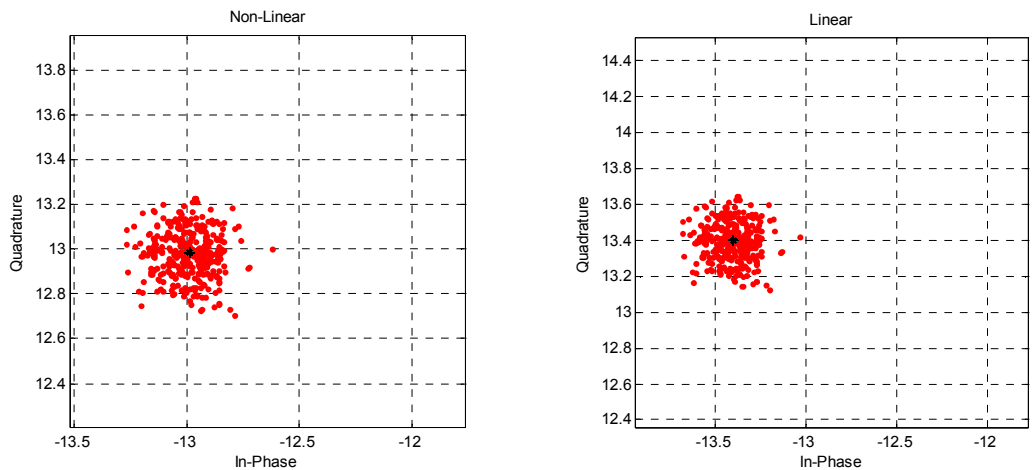

Fig.10: Point in the scatter plot at top left corner.

From the above plots we can see that the effect at high energy levels is high.

## 5. CONCLUSIONS

To have a high speed data rate, we require Multilevel Modulation schemes. And from the above observations we can conclude that effects of non linearities in RF Amplifiers cause severe performance losses for these modulation schemes. Hence a proper study of these effects will result in obtaining the optimum tradeoff between data rate and power efficiency.

## ACKNOWLEDGEMENTS

We would like to thank Dr.Shirshu Verma for his guidance throughout the compilation of the paper. We would like to thank Prof. Peter Jung and Dr. Alexander Viessmann for their valuable guidance throughout the project.